\documentclass[aps, prd, reprint, showpacs]{revtex4-1}

\usepackage{amsmath}
\usepackage{amsfonts}
\usepackage{amssymb}
\usepackage{hyperref}
\usepackage[T1]{fontenc}

\newcommand{\arxiv}[1]{\href{https://arxiv.org/abs/#1}{arXiv:#1}}

\newcommand{\bibxp}[4]{#1, {\it#2}, #3, \arxiv{#4}.}
\newcommand{\bibp}[3]{#1, {\it#2}, #3.}

\newcommand{\JourRef}[4]{{\it#1} {\bf#2} (#3) #4}
\newcommand{\BookRef}[2]{#1 (#2)}

\begin{document}

\title{Twisted statistics and the structure of Lie-deformed Minkowski spaces}

\date{\today}

\author{D.~Meljanac}\email{Daniel.Meljanac@irb.hr}
\affiliation{Division of Materials Physics, Ru\dj{}er Bo\v{s}kovi\'c Institute, Bijeni\v{c}ka~c.54, HR-10002~Zagreb, Croatia}
\author{S.~Meljanac}\email{meljanac@irb.hr}
\author{D.~Pikuti\'c}\email{dpikutic@irb.hr}
\affiliation{Division of Theoretical Physics, Ru\dj{}er Bo\v{s}kovi\'c Institute, Bijeni\v{c}ka~c.54, HR-10002~Zagreb, Croatia}
\author{Kumar~S.~Gupta}\email{kumars.gupta@saha.ac.in}
\affiliation{Saha Institute of Nuclear Physics, Theory Division, 1/AF Bidhannagar, Kolkata 700064, India}



\begin{abstract}We show that the realizations of noncommutative coordinates that are linear in the Lorentz generators form a closed Lie algebra under certain conditions. The star product and the coproduct for the momentum generators are obtained for these Lie algebras and the corresponding twist satisfies the cocycle and normalization conditions. We also obtain the twisted flip operator and the $\mathcal R$-matrix that define the statistics of particles or quantum fields propagating in these noncommutative spacetimes. The Lie algebra obtained in this work contains a special case which has been used in the literature to put bounds on noncommutative parameters from the experimental limits on Pauli forbidden transitions. The general covariant framework presented here is suitable for analyzing the properties of particles or quantum fields at the Planck scale.\end{abstract}



\maketitle


\section{Introduction}

Noncommutative geometry is one of the candidates for describing physics at the Planck scale. A combined analysis of Einstein's general relativity and Heisenberg's uncertainty principle leads to a very general class of noncommutative spacetimes \cite{dfr-1,dfr-2}, examples of which include the Groenewald-Moyal plane \cite{majid} and the $\kappa$-Minkowski algebra \cite{luru-1,luru-2,luru-3, amelino-1,amelino-2, pachol-strajn, borowiec-gupta}. Analysis of quantum mechanics and field theory on such noncommutative space-times provides a glimpse of the physics at the Planck scale, in which the spin-statistics \cite{PCT} theorem plays a particularly important role. The statistics of particles or quantum fields is defined by observing how a two particle state behaves under exchange, which is implemented by the flip operator. In order that the particle statistics is preserved under symmetry operations in the noncommutative spaces \cite{chai1,chai2}, the usual flip operator must be replaced with a twisted flip operator \cite{chaks,b1,trg1}. This result has profound consequences for the spin-statistics theorem and  the Pauli exclusion principle in noncommutative spaces, as it gives rise to certain Pauli forbidden transitions. The experimental data on such forbidden transitions in turn put bounds on the parameters of the noncommutative space-times \cite{chaks,balachandran,balachandran2}.

The twisted flip operator also plays a crucial role  in the development of quantum field theories in noncommutative spaces. The usual bosonic or fermionic statistics of a quantum field is encoded in the oscillator algebra of the creation and annihilation operators. In a noncommutative space-time, the usual oscillator algebra changes to a twisted one, which leads to novel physical effects for field theories and gravity \cite{bgrav,bgauge,ncfields}, as well as for quantum fields around noncommutative black holes \cite{ncbh}. 

It is known \cite{majid, drinfeld2-1,drinfeld2-2,drinfeld2-3} that deformations of a symmetry group can be realized through application of Drinfeld twists on that symmetry group \cite{drinfeld1, drinfeld3, EPJC11, aschieri}. The main virtue of the twist formulation is that the deformed, twisted symmetry algebra is the same as the original undeformed one. There is only a change in the coalgebra structure \cite{majid} which then leads to the same single particle Hilbert space and free field structure as in the corresponding commutative theory.

In this work, instead of starting with a given algebra, we begin with the realizations of noncommutative coordinates which are linear in Lorentz generators and demand that the noncommutative coordinates close a Lie algebra. Next we proceed to find the star products and the coproducts of the momentum generators and obtain the corresponding twists from those coproducts. These are Drinfeld twists satisfying normalization and cocycle conditions. They generate a new twisted Poincar\'e Hopf algebra. In a special case we obtain $\kappa$-Poincar\'e Hopf algebra only for the light-like deformation \cite{IJMPA, JHEP,light-like-kappa,EPJC}. Recently, a new type of noncommutative algebra has been used to analyze Pauli forbidden transitions \cite{balachandran,balachandran2}. In this paper we find a covariant generalization of that algebra, construct the flip operator and analyze the corresponding twisted statistics. 

The paper is organized as follows. In Section 2, we present realizations of noncommutative coordinates which are linear in Lorentz generators and demand that they close a Lie algebra. In Section 3, star products and coproducts of momentum generators are obtained. In Section 4, the corresponding Drinfeld twists are given. They generate twisted Poincar\'e Hopf algebra. Flip operator and twisted statistics are presented in Section 5. We conclude the paper in Section 6 with discussions and an outlook.

\section{Realizations of noncommutative coordinates linear in Lorentz generators}
Let us define the undeformed phase space, namely the Heisenberg algebra $\mathcal H$, generated by coordinates $x^\mu$ and momenta $p_\mu$, $\mu = 0,1, ... ,n-1$:
\begin{equation}\begin{split}
[x^\mu, x^\nu] &=0, \\
[p_\mu, p_\nu] &=0, \\
[p^\mu, x_\nu] &= -i\delta^\mu_\nu.
\end{split}\end{equation}

The Poincar\'e algebra  in Minkowski space ${\cal M}_{1,n-1}$ is generated by the Lorentz generators $M_{\mu\nu}$ and the momentum generators $p_\mu$, satisfying the relations
\begin{equation}\begin{split} 
[M_{\mu\nu}, M_{\rho\sigma}]&=i(\eta_{\mu\rho}M_{\nu\sigma} - \eta_{\nu\rho}M_{\mu\sigma} \\
&\phantom{={}}
-\eta_{\mu\sigma}M_{\nu\rho} + \eta_{\nu\sigma}M_{\mu\rho}), \\
[M_{\mu\nu}, p_\lambda]&=i(\eta_{\mu\lambda} p_\nu -  \eta_{\nu\lambda}p_\mu), \\
[p_\mu, p_\nu]&=0,
\end{split}\end{equation}
where $\eta_{\mu\nu}=diag(-1,1,...,1)$. The Lorentz generators $M_{\mu\nu}$ are represented by
\begin{equation}\begin{split}
M_{\mu\nu}&=x_\mu p_\nu - x_\nu p_\mu, \\
M_{\mu\nu}&=-M_{\nu\mu}, \\
M^\dagger_{\mu\nu}&=M_{\mu\nu},
\end{split}\end{equation}
and satisfy the relation
\begin{equation}
[M_{\mu\nu}, x_\lambda]=i(\eta_{\mu\lambda} x_\nu -  \eta_{\nu\lambda}x_\mu). 
\end{equation}

The most general noncommutative coordinates $\hat x_\mu$ linear in the Lorentz generators are
\begin{equation}\begin{split}
\hat x_\mu &= x_\mu + \frac l2 K_\mu{}^{\rho\sigma} M_{\rho\sigma}, \qquad K_\mu{}^{\rho\sigma}\in\mathbb R \\
\hat x^\dagger_\mu &= \hat x_\mu,
\end{split}\end{equation}
where $l$ is a dimensionful constant whose value is typically of the order of the Planck length.

Commutation relations $[\hat x_\mu, \hat x_\nu]$ are then linear in $\hat x_\lambda$ and $M_{\alpha\beta}$. We would like to determine the conditions on $K_\mu{}^{\rho\sigma}$ under which the  commutation relations $[\hat x_\mu, \hat x_\nu]$ would generate Lie algebra closed in $\hat x_\lambda$. The conditions that we have found leads to two types of solutions, which are discussed below. 

The solutions for the type $i)$ are given by
\begin{equation}\label{casei}
 \hat x_\mu=x_\mu + lu^\alpha M_{\alpha\mu},
\end{equation}
where $u^\alpha\in\mathcal M_{1,n-1}$ is a light-like vector $u^2=0=u^\alpha u_\alpha$. The corresponding noncommutative space is the $\kappa$-Minkowski space \cite{IJMPA, JHEP, light-like-kappa, EPJC} defined by 
\begin{equation}\label{xLiei}
[\hat x^\mu, \hat x^\nu]=il(u^\mu\hat x^\nu - u^\nu \hat x^\mu).
\end{equation}

The solutions for the type $ii)$ are given by
\begin{equation}\label{caseii}
 \hat x_\mu = x_\mu + a_\mu \frac l2 \theta^{\alpha\beta} M_{\alpha\beta},
\end{equation}
where $\theta^{\mu\nu}=-\theta^{\mu\nu}$, $a_\mu\theta^{\mu\nu}=0$ and $a_\mu \in \mathcal M_{1,n-1}$, $a^2\in \{-1,0,1\}$. Then, the noncommutative coordinates $\hat x_\mu$ close a Lie algebra
\begin{equation}\label{xLieii}
[\hat x_\mu, \hat x_\nu]=il(a_\mu \theta_{\nu\alpha} - a_\nu \theta_{\mu\alpha})\hat x^\alpha
\end{equation}
and the generators $\hat x_\mu$, $M_{\mu\nu}$ also close a Lie algebra
\begin{equation}\begin{split}
[M_{\mu\nu}, \hat x_\rho]&=i(\hat x_\mu \eta_{\nu\rho} - \hat x_\nu \eta_{\mu\rho}) \\
&\phantom{={}} + ia_\rho l(M_{\mu\alpha}\theta^\alpha{ }_\nu - M_{\nu\alpha} \theta^\alpha{}_\mu).
\end{split}\end{equation}
The momentum generators $p_\mu$ and the noncommutative coordinates $\hat x_\mu$ form a deformed Heisenberg algebra given by equation \eqref{xLieii} and
\begin{equation}
[p_\mu, \hat x_\nu] = -i\left(
\eta_{\mu\nu}+a_\nu l \theta_{\mu\alpha}p^\alpha
\right).
\end{equation}

We point out that the Casimir operator is $\mathcal C = p^2=p_\alpha p^\alpha$ and the dispersion relation is thus undeformed.

\subsection{Special cases of type $ii)$}
A special case of the type $ii)$ solutions obtained above is given in \cite{balachandran,balachandran2}, where the following algebra was considered in 3+1 dimensions
\begin{equation}\begin{split}\label{xxb1}
[\hat x_0, \hat x_i] &= i \chi \epsilon_{ijk} n_k \hat x_j, \\
[\hat x_i, \hat x_j]&=0.
\end{split}\end{equation}
It was shown in \cite{balachandran,balachandran2} that the algebra (\ref{xxb1}) leads to Pauli violating transitions. The experimental limits on such transitions were then used to put bounds on the noncommutative parameters.

The algebra (\ref{xxb1}) can be written covariantly as
\begin{equation}\label{xxb2}
[\hat x_\mu, \hat x_\nu] = il(a_\mu \epsilon_{\alpha\beta\gamma\nu} - a_\nu \epsilon_{\alpha\beta\gamma\mu})
a^\alpha b^\beta \hat x^\gamma.
\end{equation}
This Lie algebra reduces to case given in \cite{balachandran,balachandran2} for $\chi=l$, $a_\mu=(1,\vec0)$ and $b_\mu=(0,\vec n)$, where $a_\mu, b_\mu \in \mathcal M_{1,3}$ and $\theta_{\mu\nu}=\epsilon_{\mu\nu\alpha\beta}a^\alpha b^\beta$. Linear realization for Lie algebra \eqref{xxb2} is
\begin{equation}
\hat x_\mu = x_\mu + \frac l2 a_\mu \epsilon_{\alpha\beta\gamma\delta}a^\alpha b^\beta M^{\gamma\delta}.
\end{equation}
Note that realization which follows from \cite{balachandran,balachandran2} is non-linear in momentum generators.

Another special case of \eqref{xxb2} is given in 3+1 dimensions \cite{lukierski06plb} with
\begin{equation}\begin{split} 
&a_\mu=\zeta_\mu, \qquad \theta_{\mu\nu}=\frac{\alpha_\mu \beta_\nu - \alpha_\nu \beta_\mu}2, \qquad \\
& \alpha\cdot\beta=0, \qquad |\alpha^2|=|\beta^2|=1.
\end{split}\end{equation}
Realization which follows from \cite{lukierski06plb} is nonlinear in momentum generators.

A third special case is $\rho$-deformed Minkowski space in 2+1 dimensions, discussed in \cite{ro}.

\subsection{Exterior derivatives and one-forms}

For realizations linear in the Lorentz generators, it is easy to define the corresponding Lorentz one-forms $\hat\xi^\mu=[d,\hat x^\mu]$, where $d$ is the exterior derivative \cite{IJMPA, JHEP}
\begin{equation}
d=i\xi^\alpha p_\alpha
\end{equation}
given in terms of the undeformed super Heisenberg algebra
\begin{equation}
\{\xi^\mu, \xi^\nu \} = 0, 
\qquad \{\mathfrak q_\mu, \mathfrak q_\nu\} = 0, 
\qquad \{\xi^\mu, \mathfrak q_\nu \} = \delta^\mu_\nu.
\end{equation}
In this case $\{\hat\xi^\mu, \hat\xi^\nu\}=0$, $d^2=0$ and $\hat x^\mu$ and $\hat\xi^\mu$ generate a Lie superalgebra where $[\hat\xi^\mu, \hat x^\nu]=iK^{\mu\nu}{}_\lambda \hat \xi^\lambda$ is closed in $\hat\xi^\lambda$. In this case, the Lorentz generators are given by
\begin{equation}
M_{\mu\nu}=x_\mu p_\nu - x_\nu p_\mu - i(\xi_\mu \mathfrak  q_\nu - \xi_\nu \mathfrak  q_\mu).
\end{equation}



\section{Star product and coproduct of momentum generators}
Action $\triangleright:\mathcal H \otimes \mathcal A \to \mathcal A$ is defined by
\begin{equation}\begin{split}
x^\mu\triangleright f(x) &= x^\mu f(x),\\
p_\mu\triangleright f(x) &=-i\frac{\partial f}{\partial x^\mu}.
\end{split}\end{equation}
Then, for realizations \eqref{casei} and \eqref{caseii}, we have
\begin{align}
e^{ik\cdot \hat x} \triangleright e^{iq\cdot x}&= e^{i\mathcal P(k,q)\cdot x}, \\
e^{ik\cdot \hat x}\triangleright 1 &=e^{iK(k)\cdot x}, \\
e^{iK^{-1}(k)\cdot \hat x}\triangleright 1 &=e^{ik\cdot x},
\end{align}
where $k,q\in\mathcal M_{1,n-1}$, $\mathcal P_\mu(k,0)=K_\mu(k)$, $\mathcal P_\mu(0,q)=q_\mu$ and $K^{-1}_\mu(k)$ is the inverse map of $K_\mu: \mathcal M_{1,n-1} \to \mathcal M_{1,n-1}$, in the sense that $K_\mu(K^{-1}(k))=K^{-1}_\mu(K(k))=k_\mu$.

The star product is defined by
\begin{equation}\begin{split} 
e^{ik\cdot x} \star e^{iq\cdot x} &= e^{iK^{-1}(k)\cdot \hat x}\triangleright e^{iq\cdot x} \\
&=e^{i\mathcal P(K^{-1}(k),q)\cdot x}=e^{i\mathcal D(k,q)\cdot x}.
\end{split}\end{equation}
For Lie-algebraic deformations of Minkowski space, the corresponding star product is associative and is defined by $\mathcal D_\mu(k,q)$. Deformed addition of momenta is given by
\begin{equation}
(k\oplus q)_\mu = \mathcal D_\mu(k,q).
\end{equation}
For the light-like $\kappa$ deformation of Minkowski space \eqref{casei}, results for $\mathcal P_\mu(k,q)$ and $\mathcal D_\mu(k,q)$ are given in \cite{EPJC}.


For Lie-algebraic deformations of Minkowski space of type $ii)$ \eqref{xLieii} with realizations linear in Lorentz generators \eqref{caseii}, the results are
\begin{align}
K_\mu(k)&=\left(\frac{e^{(a\cdot k)l\theta}-1}{(a\cdot k)l\theta} \right){}^\alpha{}_\mu k_\alpha, \\
K^{-1}_\mu(k)&=\left(\frac{(a\cdot k)l\theta}{e^{(a\cdot k)l\theta}-1} \right){}^\alpha{}_\mu k_\alpha, \\
\mathcal P_\mu(k,q)&=\left(\frac{e^{(a\cdot k)l\theta}-1}{(a\cdot k)l\theta} \right){}^\alpha{}_\mu k_\alpha
+ (e^{(a\cdot k)l\theta})^\alpha{}_\mu q_\alpha, \\
\mathcal D_\mu(k,q)&=k_\mu +  (e^{(a\cdot k)l\theta})^\alpha{}_\mu q_\alpha.
\end{align} 
Note that we can introduce a new momentum generator
\begin{equation}
p^W_\mu=K^{-1}_\mu(p),
\end{equation}
which corresponds to Weyl-symmetric realization\cite{universal-formula, sasa}, with the property
\begin{equation}
p^W_\mu e^{ik\cdot \hat x} \triangleright 1 =
p^W_\mu \triangleright e^{iK(k)\cdot x} =
 k_\mu e^{iK(k)\cdot x}.
\end{equation}

For linear realizations of noncommutative coordinates $\hat x_\mu = x_\mu + K_{\beta\mu\alpha}x^\alpha p^\beta$, the momenta $p_\mu$ and $p^W_\mu$ are related by \cite{EPJC}
\begin{equation}
p_\mu = \left( \frac{1-e^{-\mathcal K}}{\mathcal K}\right){^\alpha}_\mu p^W_\alpha,
\end{equation}
where $\mathcal K_{\mu\nu} = -K_{\nu\alpha\mu}(p^W)^\alpha$.

For realization \eqref{caseii}, $\mathcal K_{\mu\nu}$ is
\begin{equation}
\mathcal K_{\mu\nu}=-(a\cdot p^W)l \theta_{\mu\nu}.
\end{equation}
Since $a^\alpha \theta_{\alpha\mu}=0$, it follows that $a^\alpha(\mathcal K^n)_{\alpha\mu}=0$ and therefore,
\begin{equation}
a\cdot p^W = a\cdot p
\end{equation}
and $\mathcal K_{\mu\nu}$ is also given by
\begin{equation}
\mathcal K_{\mu\nu}=-(a\cdot p)l \theta_{\mu\nu}.
\end{equation}


Coproduct of momentum generator $p_\mu$ is given by
\begin{equation}
\Delta p_\mu = \mathcal D_\mu(p\otimes 1, 1\otimes p)=p_\mu\otimes 1 + (e^{-\mathcal K})_{\alpha\mu}\otimes p^\alpha.
\end{equation}




\section{Twist from realization linear in Lorentz generators}

Result for the twist $\mathcal F$, constructed from realization of noncommutative coordinates $\hat x_\mu$ linear in generators of $\mathfrak{gl}(n)$, is presented in \cite{EPJC} (see also \cite{IJMPA, JHEP, IJMPA-twist, PLB2017} and is given by
\begin{equation}
\mathcal F=\exp(-ip^W_\alpha \otimes(\hat x^\alpha -x^\alpha)).
\end{equation}
This result holds for $\hat x^\alpha$ given in \eqref{casei} and \eqref{caseii}.

For light-like $\kappa$-deformed Minkowski space \eqref{xLiei} and linear realization \eqref{casei} of type $i)$, twist is constructed and discussed in \cite{IJMPA, JHEP, light-like-kappa, EPJC}, with 
\begin{equation}
p^W_\mu=\left(p_\mu + \frac{a_\mu l}2p^2 \right)\frac{\ln(1+(a\cdot p)l)}{(a\cdot p)l}.
\end{equation}

For Lie-algebraic deformation of Minkowski space \eqref{xLieii} and linear realizations \eqref{caseii} of type $ii)$, twist is given by
\begin{equation}\label{twist1}
\mathcal F=\exp\left(-i\frac{a\cdot p}{2}l\otimes \theta^{\alpha\beta}M_{\alpha\beta} \right).
\end{equation}
It is an Abelian twist and it satisfies normalization and cocycle condition, see \cite{EPJC}. From this twist \eqref{twist1}, we find that
\begin{align}
\begin{split} 
\hat x_\mu &= m\left[\mathcal F^{-1}(\triangleright\otimes1)(x_\mu\otimes1) \right] \\
&=x_\mu+a_\mu\frac l2 \theta^{\alpha\beta}M_{\alpha\beta}, 
\end{split}\\
e^{iK^{-1}(k)\cdot\hat x}&=m\left[\mathcal F^{-1}(\triangleright\otimes1)(e^{ik\cdot x}\otimes1) \right], \\
e^{iK^{-1}(k)\cdot\hat x}\triangleright1&=e^{ik\cdot x}, \\
\begin{split} 
e^{ik\cdot x}\star e^{iq\cdot x}&=m\left[ 
\mathcal F^{-1}(\triangleright\otimes\triangleright)(e^{ik\cdot x}\otimes e^{iq\cdot x})
\right]\\&=e^{i\mathcal D(k,q)\cdot x},\end{split}
\end{align}
where $m$ is the multiplication map $m: \mathcal H \otimes \mathcal H \to \mathcal H$. These results represent consistency check of the construction \cite{EPJC, PLB2017}.

\subsection{Twisted Poincar\'e Hopf algebra}
Using twist $\mathcal F$ \eqref{twist1}, the twisted Poincar\'e Hopf algebra is easily obtained. Coproducts of momentum $p_\mu$ and Lorentz generators $M_{\mu\nu}$ are
\begin{align}
\Delta p_\mu &= \mathcal F\Delta_0 p_\mu \mathcal F^{-1} =
p_\mu\otimes 1 + {(e^{\mathcal K})_\mu}^\alpha\otimes p_\alpha,\\
\begin{split}
\Delta M_{\mu\nu}&= \mathcal F\Delta_0 M_{\mu\nu} \mathcal F^{-1} \\&=
M_{\mu\nu}\otimes1
+{(e^{\mathcal K})_\mu}^\alpha {(e^{\mathcal K})_\nu}^\beta \otimes M_{\alpha\beta}\\
&\phantom{={}}+\frac12(a_\mu p_\nu - a_\nu p_\mu) \otimes \theta^{\alpha\beta}M_{\alpha\beta},
\end{split}
\end{align}
where $\Delta_0 p_\mu = p_\mu \otimes 1 + 1 \otimes p_\mu$ and $\Delta_0 M_{\mu\nu}=M_{\mu\nu} \otimes 1 + 1 \otimes M_{\mu\nu}$.

The antipodes are
\begin{align}
S(p_\mu)&=-{(e^{\mathcal K})_\mu}^\alpha p_\alpha,\\
\begin{split} 
S(M_{\mu\nu})&=-{(e^{-\mathcal K})_\mu}^\alpha {(e^{-\mathcal K})_\nu}^\beta M_{\alpha\beta}\\
&\phantom{={}}-\frac12(a_\mu S(p_\nu) - a_\nu S(p_\mu))  \theta^{\alpha\beta}M_{\alpha\beta}.
\end{split}
\end{align}
The counit is trivial
\begin{equation}
\epsilon(p_\mu)=0, \qquad \epsilon(M_{\mu\nu})=0, \qquad \epsilon(1)=1.
\end{equation}

\subsection{$\mathcal R$-matrix} 
The $\mathcal R$-matrix is given by
\begin{equation}\begin{split}\label{Rmatrix} 
\mathcal R &= \tilde{\mathcal F} \mathcal F^{-1}\\&=
\exp \left[ \frac{il}2\theta^{\alpha\beta} \left(
a\cdot p \otimes M_{\alpha\beta} - M_{\alpha\beta} \otimes a\cdot p
\right)\right] \\
&=e^r = 1\otimes1 + r + \mathcal O(l^2),
\end{split}\end{equation}
where $\tilde{\mathcal F}=\tau_0 \mathcal F \tau_0$ is transposed twist, $\tau_0: \mathcal H \otimes \mathcal H \to \mathcal H \otimes \mathcal H$ is a linear map such that $\tau_0(A\otimes B) = B \otimes A ~~ \forall A, B \in \mathcal H$ and $r$ is the classical $r$ matrix.

Lie algebraic deformations of Minkowski space have infinitely many realizations of $\hat x'_\mu$ in terms of undeformed Heisenberg algebras, generated by $x'_\mu$ and $p'_\mu$, which are related by similarity transformations, see for example \cite{m-s-1,m-s-2,m-s-3}. Large class of realizations $\hat x'_\mu$ and $p'_\mu$ (non-linear in generators) of the deformed Minkowski space \eqref{xLieii} leads to Abelian twists $\mathcal F'$, which give the same form of $\mathcal R$ matrix \eqref{Rmatrix}. Note that $\Delta_0 = \mathcal F^{-1}\Delta \mathcal F = \tilde{\mathcal F}^{-1}\tilde\Delta\tilde{\mathcal F}$ 
and $\mathcal R\Delta = \tilde\Delta\mathcal R$.

\section{Flip operator and twisted statistics}

As mentioned before, in noncommutative spacetimes, the flip operator that defines the particles statistics has to be replaced with the twisted flip operator. The twisted flip operator $\tau$ is defined by
\begin{equation}
\tau=\mathcal F \tau_0 \mathcal F^{-1}=\tau_0 \mathcal R
\end{equation}
and it satisfies the conditions
\begin{align}
[\Delta h, \tau ] &= 0, \quad \forall h\in\mathcal U(\mathcal P),\\
\tau^2&=1\otimes1
\end{align}

The projectors for the twisted symmetric and anti-symmetric sectors of the Hilbert space are given by $\frac12(1\otimes1\pm\tau)$. We define deformed bosonic state as
\begin{equation}\label{bstate}
f\otimes g = \tau(f\otimes g),
\end{equation}
or equivalently 
\begin{equation}
\mathcal F^{-1}(\triangleright\otimes\triangleright)(f\otimes g) = \tilde{\mathcal F}^{-1}(\triangleright\otimes\triangleright)(g\otimes f).
\end{equation}
Now, defining the twisted tensor product
\begin{equation}
f\otimes_{\mathcal F} g = \mathcal F^{-1}(\triangleright\otimes\triangleright)(f\otimes g),
\end{equation}
we get
\begin{equation}
f\otimes_{\mathcal F}g=\tau_0(f\otimes_{\mathcal F}g)
\end{equation}
For the product of two bosonic fields $\phi(x)$ and $\phi(y)$ under interchange, additional factor appears compared to commutative case. This can be calculated using \eqref{bstate} and one gets
\begin{equation}
\mathcal R(\triangleright\otimes\triangleright)(\phi(x)\otimes\phi(y))=\phi(y)\otimes\phi(x)
\end{equation}
Expressing $\phi$ in the above equation using Fourier transforms and using twisted flip operator or equivalently the $\mathcal R$-matrix in momentum space, we are led to deformed commutation relations between annihilation operators. These twisted creation and annihilation operators should be used to perform any calculations in the corresponding quantum field theories \cite{bgrav,bgauge,ncfields}.

\section{Outlook and discussion}

In this paper we have started with the realizations of noncommutative coordinates which are linear in Lorentz generators and have obtained the conditions under which they form a closed Lie algebra. We have obtained the star products and the coproducts of the momentum generators and have written down the corresponding twist operator. This has been shown to be a Drinfeld twist which satisfies the normalization and the cocycle conditions. 

As a special case of our results, we have obtained the covariant generalization of an algebra which was used in \cite{balachandran,balachandran2} to put bounds on the noncommutative parameters from the experimental data on Pauli forbidden transitions. Thus the results presented here open up the possibility of a more general analysis of the Pauli forbidden transitions within the context of the noncommutative framework.

Using the Drinfeld twists, we have obtained the twisted flip operators which are the basic building blocks of constructing symmetric and anti-symmetric sectors of the Hilbert space in the quantum theory. The new twisted flip operators presented here can be used to construct and analyze physical processes within the framework quantum field theories. Since noncommutativity is expected to be a feature at the Planck scale, the predictions from such twisted quantum field theories would provide a glimpse of the physics at the Planck scale. In particular, the quantum field theories in the near-horizon region of black holes are expected to carry signatures of the quantum gravity scale. The analysis presented here can be used to model such field theories which can provide a hint about the space-time structure at the Planck scale.  

\section*{Acknowledgements}
The work by S.~M. and D.~P. has been supported by Croatian Science Foundation under the Project No. IP-2014-09-9582 as well as by the H2020 Twinning project No. 692194, ``RBI-T-WINNING''.

\end{document}